\documentclass[]{tCPH2e}

\begin{document}
\doi{10.1080/0010751YYxxxxxxxx}
 \issn{1366-5812}
\issnp{0010-7514}

\jvol{00} \jnum{00} \jyear{2011} 

\markboth{Gomboc}{Contemporary Physics}

\title{{\itshape Unveiling the Secrets of Gamma Ray Bursts}}

\author{Andreja Gomboc$^{a,b}$$^{\ast}$\thanks{$^\ast$Corresponding author. Email: andreja.gomboc@fmf.uni-lj.si
\vspace{6pt}} \\\vspace{6pt} $^{a}${\em{Faculty of Mathematics and Physics, University of Ljubljana, Slovenia}}; \\
$^{b}${\em{Centre of Excellence SPACE-SI, Ljubljana, Slovenia}}}

\maketitle

\begin{abstract}
Gamma Ray Bursts (GRBs) are unpredictable and brief flashes of gamma rays that occur about once a day in random locations in the sky. Since gamma rays do not penetrate the Earth's atmosphere, they are detected by satellites, which automatically trigger ground-based telescopes for follow-up observations at longer wavelengths. In this introduction to Gamma Ray Bursts we review how building a multi-wavelength picture of these events has revealed that they are the most energetic explosions since the Big Bang and are connected with stellar deaths in other galaxies. However, in spite of exceptional observational and theoretical progress in the last 15 years, recent observations raise many questions which challenge our understanding of these elusive phenomena. Gamma Ray Bursts therefore remain one of the hottest topics in modern astrophysics.

\bigskip

\bigskip
\bigskip

\end{abstract}

\section{Introduction}
Gamma Ray Bursts (GRBs) are at the intersection of many different areas of astrophysics: they are relativistic events connected with the end stages of stars; they reveal properties of their surrounding medium and of their host galaxies; they emit radiation from gamma-rays to radio wavelengths, as well as possibly non-electromagnetic signals, such as neutrinos, cosmic rays and gravitational waves. Due to their enormous luminosities, they can be detected even if they occur at vast distances, and are therefore also of great interest for cosmology. 

Let us first briefly review some basic properties of GRBs. They are unpredictable and non-repetitive violent flashes of gamma rays coming from random directions in the sky at random times and lasting from $\sim 0.01\, $s to $\approx 1000\, $s. When they occur, they outshine all other sources of gamma radiation. Their gamma ray spectrum is non-thermal, with the bulk of energy emitted in 0.1 to 1 MeV range. Gamma ray emission is followed at longer wavelengths by so-called afterglows: these appear as point-like sources in X-rays, ultraviolet, optical, infrared and radio wavebands, and fade away in a few hours to weeks, sometimes months. Deeper observations usually reveal surrounding host galaxies. Spectroscopic observations of the afterglows of GRBs and host galaxies enable us to measure their cosmological redshifts $z$\footnote[1]{The wavelength of light traveling from a source to an observer through the expanding Universe increases by the factor $\lambda_{\rm observed}/\lambda_{\rm emitted}=1+z$ (i.e., $z=(\lambda_{\rm observed} - \lambda_{\rm emitted})/\lambda_{\rm emitted}= \Delta \lambda/\lambda_{\rm emitted}$). This is also the factor by which the Universe has expanded between the time of the emission and reception of light. Cosmological redshift, $z$, is therefore a measure of the Universe's size (and adopting a certain cosmological model, also the Universe's age) at the time the light left the observed astronomical object.}, and infer their distances. From observed fluxes ($F$) and known distances ($d$), assuming that GRBs emit radiation isotropically, we can estimate that the isotropic equivalent luminosity $L_{\rm iso}= {F \cdot{4\pi d^2}}$ is in the range of $L_{\rm iso}\sim 10^{40} - 10^{47}\, $W, i.e. total isotropic energy output of these events is $E_{\rm iso} \sim 10^{42} - 10^{48}\, $J, which is comparable to the rest energy of the Sun, $M_\odot c^2$ (where $M_\odot$ is the mass of the Sun).

Another important clue to the nature of GRBs comes from their gamma ray light curves (i.e. flux in gamma rays vs time), which exhibit rapid variability on a timescale of milliseconds. Since variability on a timescale $\Delta t$ can not be produced in an area which is larger than the distance light travels during this time, we can estimate the source size to be $D \leq c \Delta t \approx 300\, $km. This immediately tells us that we are dealing with a very compact stellar mass source, and given the enormous energy involved, also relativistic effects. Sources of energy which fit the bill (small, relativistic, energetic) are compact objects such as neutron stars and black holes, and their gravitational energy (with perhaps also their rotational energy providing part of the energy needed). 

There are at least two types of GRB, and according to current understanding, both are connected with end stages of stars. One type is connected with the death of massive, rapidly rotating stars, and occurs in star forming regions of blue dwarf galaxies. The other type is found in both old and star-forming galaxies; this is much less understood and is thought to be due to mergers of two compact stellar remnants (neutron stars and/or black holes).

\section{A very brief history of GRBs observational discoveries}

It is illustrative to review the more detailed characteristics of GRBs in the chronological order of their discovery \cite{zhang2004}. 

\subsection{Discovery of GRBs}
Because of the non-transparency of the Earth's atmosphere to gamma rays, the discovery of GRBs had to wait until the era of satellites. GRBs were therefore first detected in the late 1960's by USA military Vela satellites monitoring compliance with nuclear test ban treaties. It was soon realized that these flashes of gamma rays originated from outside our Solar System. Publication of their discovery several years later \cite{klebesadel}, \cite{mazets} triggered an avalanche of theoretical models for their explanation - in 1974, only one year later, there were already 15 models, and by 1992 more than a hundred. However, their nature remained a complete mystery for more than two decades, mainly because they were detectable only for tens of seconds and exclusively at gamma-ray energies, so their positions in the sky were ill-determined.

\subsection{BATSE}
The most crucial question at the time was the distance to GRBs, with direct implications for the energetics and mechanism of these events. An important step forward was the {\em Compton Gamma-Ray Observatory} (launched in 1991), and its {\em Burst and Transient Experiment} (BATSE), which recorded over 2700 GRBs, and showed that they are isotropically distributed across the sky \cite{fishman}. As there was no concentration of GRBs towards the Galactic centre or Galactic plane, this was a strong hint of GRBs' cosmological origin. However, without distance measurements, the possibility that GRBs are isotropically distributed in a halo around our Galaxy could not be ruled out.  

\begin{figure}
\begin{center}
\includegraphics[angle=0, scale=0.3]{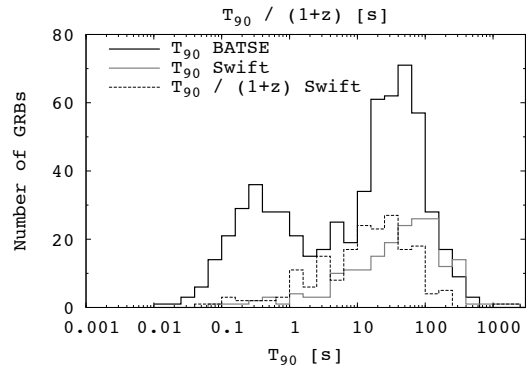}\caption{Duration $T_{90}$ of the 4B BATSE Catalog of Gamma-Ray Bursts shows two classes of GRBs (solid black line): those with $T_{90} < 2$ s are called {\em short} GRBs, and those with $T_{90}>2$ s are {\em long} GRBs. For comparison, results from more recent {\em Swift} satellite observations are shown: the grey line is the distribution of {\em Swift} bursts with known redshift over $T_{90}$ (in the observer's frame); the dashed line is the distribution of {\em Swift} bursts with known redshift over $T_{90}/(1+z)$, i.e. approximate time of duration in the GRB's rest frame. }
\label{duration}
\end{center}
\end{figure}

Another important result from BATSE was the finding that GRBs can be classified into two types according to their duration: using parameter $T_{90}$, which is defined as the duration of the time interval over which a burst emits 90\% of its fluence (fluence is the energy flux of the burst integrated over the total duration of the event), they formed a group with $T_{90}< 2\, $s, referred to as {\em short} GRBs, and a group with $T_{90}> 2\, $s, referred to as {\em long} GRBs (Fig. \ref{duration}). In addition to duration, both groups differed also in spectral hardness, i.e. the ratio of high energy photon counts to low energy photon counts was larger for short GRBs than for long GRBs. 

\begin{figure}
\begin{center}
\includegraphics[angle=0, scale=0.17]{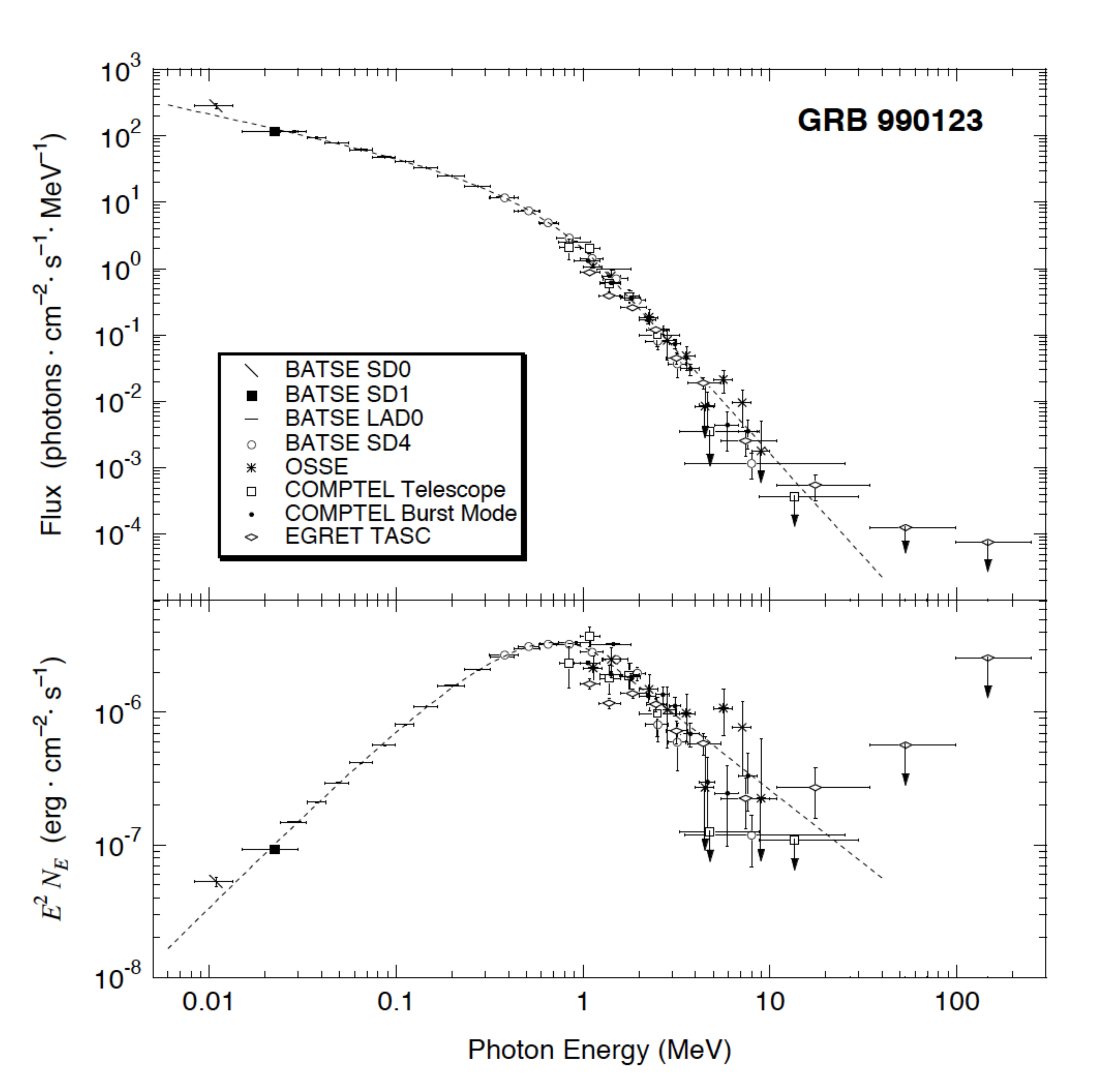}
\caption{Example of a GRB's high energy spectrum from the BATSE detectors, shown both as photon counts $N_E$ and in $E^2 N_E$ units, where $E$ is the energy of detected photons. The quantity $E^2 N_E$ tells us the amount of energy emitted in certain energy band \cite{briggs}.}
\label{GRBspec}
\end{center}
\end{figure}

BATSE observations also showed that gamma ray light curves are very diverse and difficult to classify (e.g. some have a single spiky pulse, some are smooth, with either single or multiple peaks, and others are very erratic, chaotic and spiky). Observed spectra were non-thermal, mostly rather simple, continuum spectrum well fitted by a broken power law function (Fig. \ref{GRBspec}). Typical energies of gamma ray photons emitted during a GRB were $\approx 30-1000$ keV, with the most energy emitted around 200 keV.

However, in the BATSE era, observations of GRBs remained limited to gamma-rays alone, since no follow-up observations at other wavelengths were possible. GRBs' locations by BATSE had error-boxes of a few degrees across and thus contained a large number of possible counterparts. 

\subsection{BeppoSAX and the afterglow era}

The next major breakthrough was made by the Italian-Dutch satellite {\em BeppoSAX}, which was able to produce the first quick and small error boxes of GRBs' positions in the sky (of the order of arcmin). This was crucial, since it made it realistically possible to observe a GRB's position with other, narrow field instruments several hours or days after the GRB itself.

\subsubsection{Afterglows, host galaxies and distances to GRBs}
Eight hours after the GRB~970228\footnote[2]{Numbers indicate the date of the burst: GRB~YYMMDD. If more than one burst is detected on the same day, letters A, B, C... are added to distinguish between them.} {\em BeppoSAX} detected a fading X-ray source coinciding with the GRB's position in the sky. This facilitated the detection of an optical counterpart at the same location. Such counterparts at longer wavelengths were detected also in subsequent GRBs, and were called afterglows, as they were longer-lived than gamma ray emission. They were visible for several hours to weeks and in general faded according to a power law ($F\propto t^{-\alpha}$). With the detection of optical afterglows, it was possible to pin-point GRBs' positions accurately enough to perform deeper and spectroscopic observations with large telescopes. These revealed host galaxies, and their redshifts gave solid proof that GRBs lie at cosmological distances.\footnote[3]{To be precise, at that time this was proven only for long GRBs, since due to the faintness of their optical afterglows, no optical afterglow, host galaxy or redshift of a short GRB had been observed before 2005.}

The discovery that GRBs occur in other galaxies, together with their observed brightness, led to the realization that these explosions are immense; in fact, they are the most luminous, transient objects known in the Universe and represent the most significant new astrophysical phenomenon since the discovery of quasars and pulsars. 

\subsubsection{Beamed explosions}
The next important event was the GRB~990123. Its inferred isotropic energy was $4.5\cdot 10^{47}\, $J, which is more than $M_\odot c^2$ transformed to gamma ray emission, and therefore presented a serious problem for any stellar mass model of GRBs. It was a hint that perhaps emission is not isotropic, but instead collimated. If it is collimated in two opposite jets with a half-opening angle $\theta_{\rm j} \ll 1$, this would lower the required energy budget by ${{{\theta_{\rm j}}^2}/ 2}$. On the other hand, it would also imply that, if the directions of jets are isotropic, the number of GRBs events in the Universe is in fact much larger than we can detect, since we can observe only those GRBs that accidentally point in our direction. 

\begin{figure}
\begin{center}
\includegraphics[angle=0, scale=0.25]{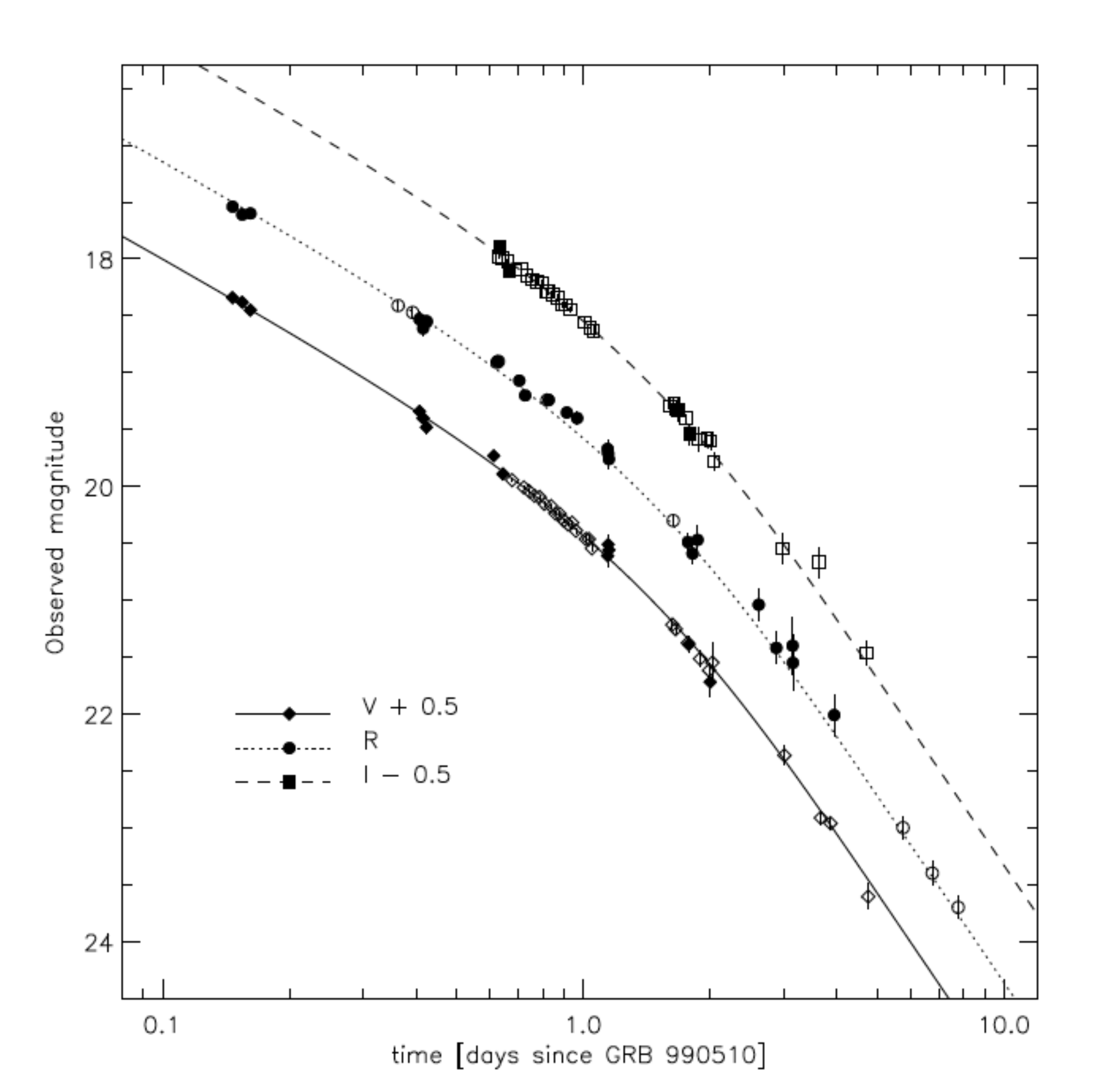}
\includegraphics[angle=0, scale=0.25]{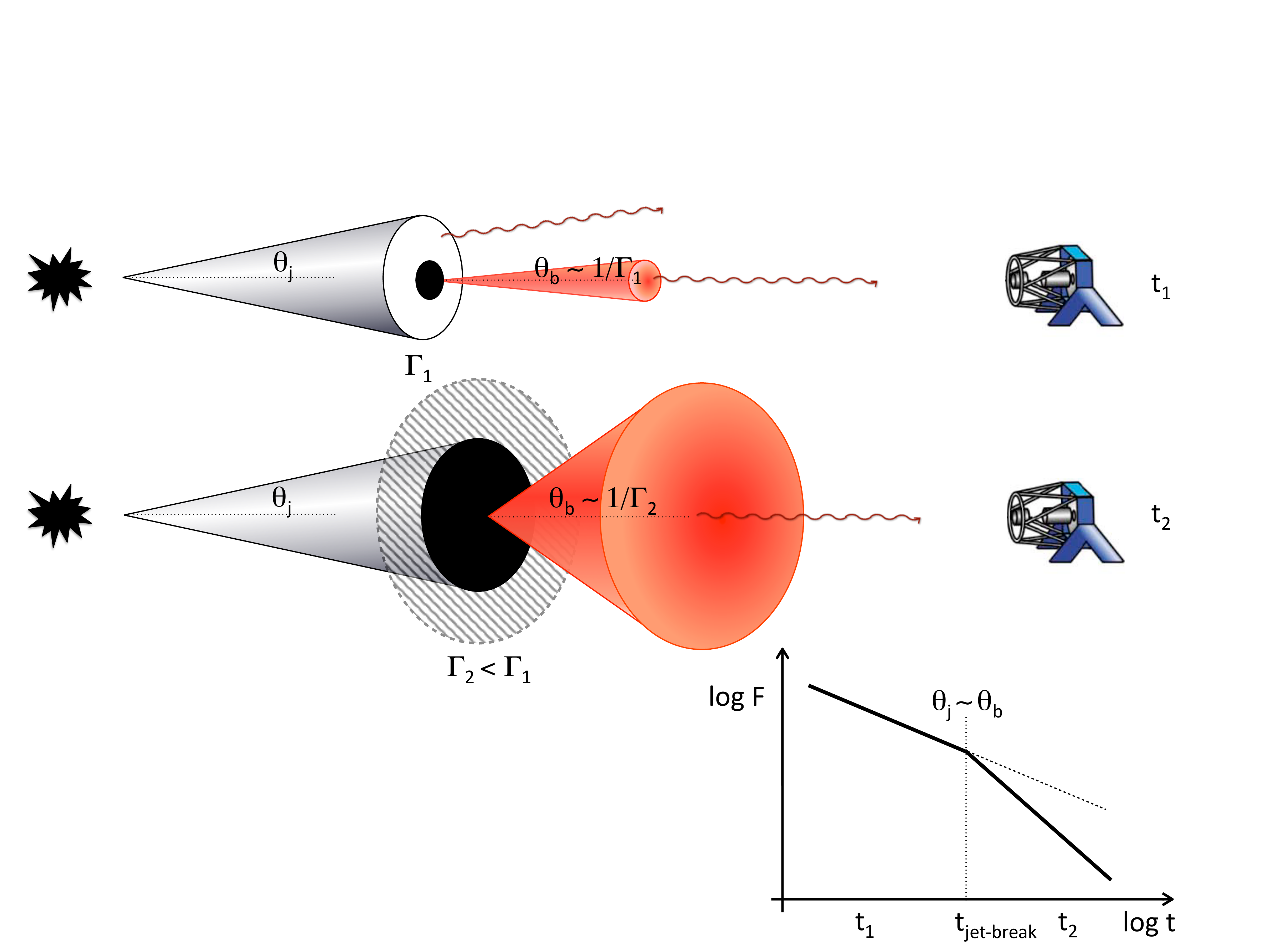}
\caption{Left: observed light curves of the optical afterglow of GRB~990510 in three filters \cite{harrison}. Right: Achromatic steepening of the light curve is expected if relativistic outflow is beamed in a narrow angle $\theta_{\rm j}$. When ejecta slows down $\theta_{\rm b}>\theta_{\rm j}$ and the observer 'misses' the emission from the dashed area, which causes the light curve to decay more steeply.}
\label{lc-break}
\end{center}
\end{figure}

Evidence for a collimated outflow came from further observations of optical afterglows' light curves (i.e. their optical brightness vs time), which showed a simultaneous break in several observational filters (Fig. \ref{lc-break} left). Such a break is expected in cases of collimated and highly relativistic moving ejecta (why the ejecta must be moving relativistically will be discussed in sec. \ref{cp}). Due to the aberration of light, the emission from material moving relativistically with Lorentz factor $\Gamma $, is strongly beamed within an angle $\theta_{\rm b} \sim {1\over {\Gamma}}$. The observer (Fig. \ref{lc-break} right) sees in his/her line of sight only a patch of ejecta inside an angle $\theta_{\rm b} $, while an emission from other areas of ejecta is beamed away. As the flow is moving through the interstellar material, it slows down, and angle $\theta_{\rm b}$ increases. As long as $\theta_{\rm b}$ is smaller than $\theta_{\rm j}$, the observer can not distinguish between spherical or collimated outflow. However, when the outflow slows down to $\theta_{\rm b} > \theta_{\rm j}$, the observer begins to see the edge of the jet cone and receives less light than in the case of a spherical outflow. This causes the light curve to decay faster. The resulting steepening in the light curve is seen simultaneously at all wavelengths emitted by the outflow, and is referred to as an achromatic break or a jet break. From the time of the break, the jet opening angle $\theta_{\rm j}$ can be deduced. Typical values for long GRBs are about 4$^{\rm o}$, and rarely exceed 10$^{\rm o}$. With available data on the beaming angles, it was estimated that there are about $\sim 100-1000$ GRBs/day occurring in the Universe.

\subsubsection{Connection between long GRBs and supernovae}
\begin{figure}
\begin{center}
\includegraphics[angle=0, scale=0.2]{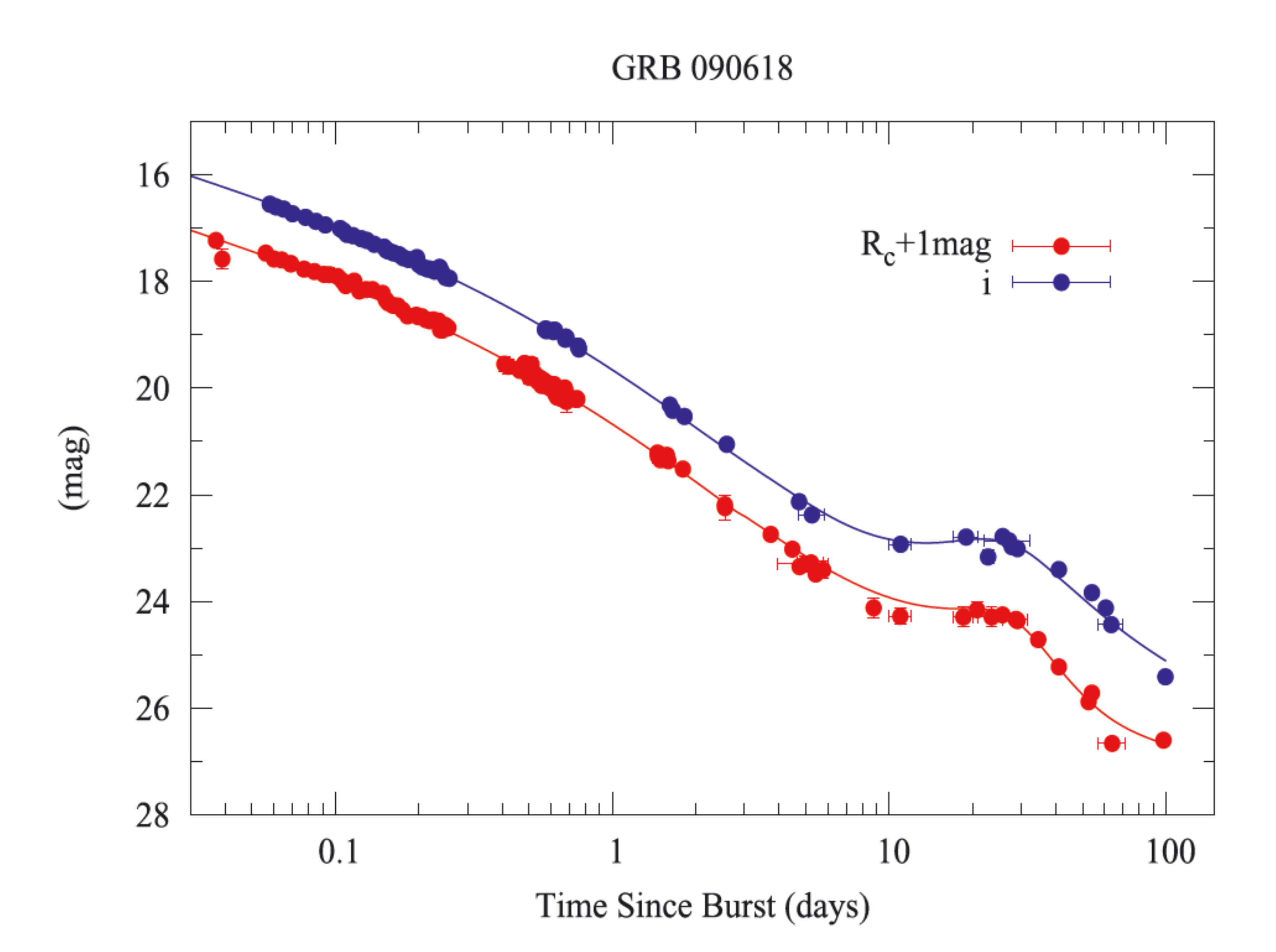}
\includegraphics[angle=0, scale=0.25]{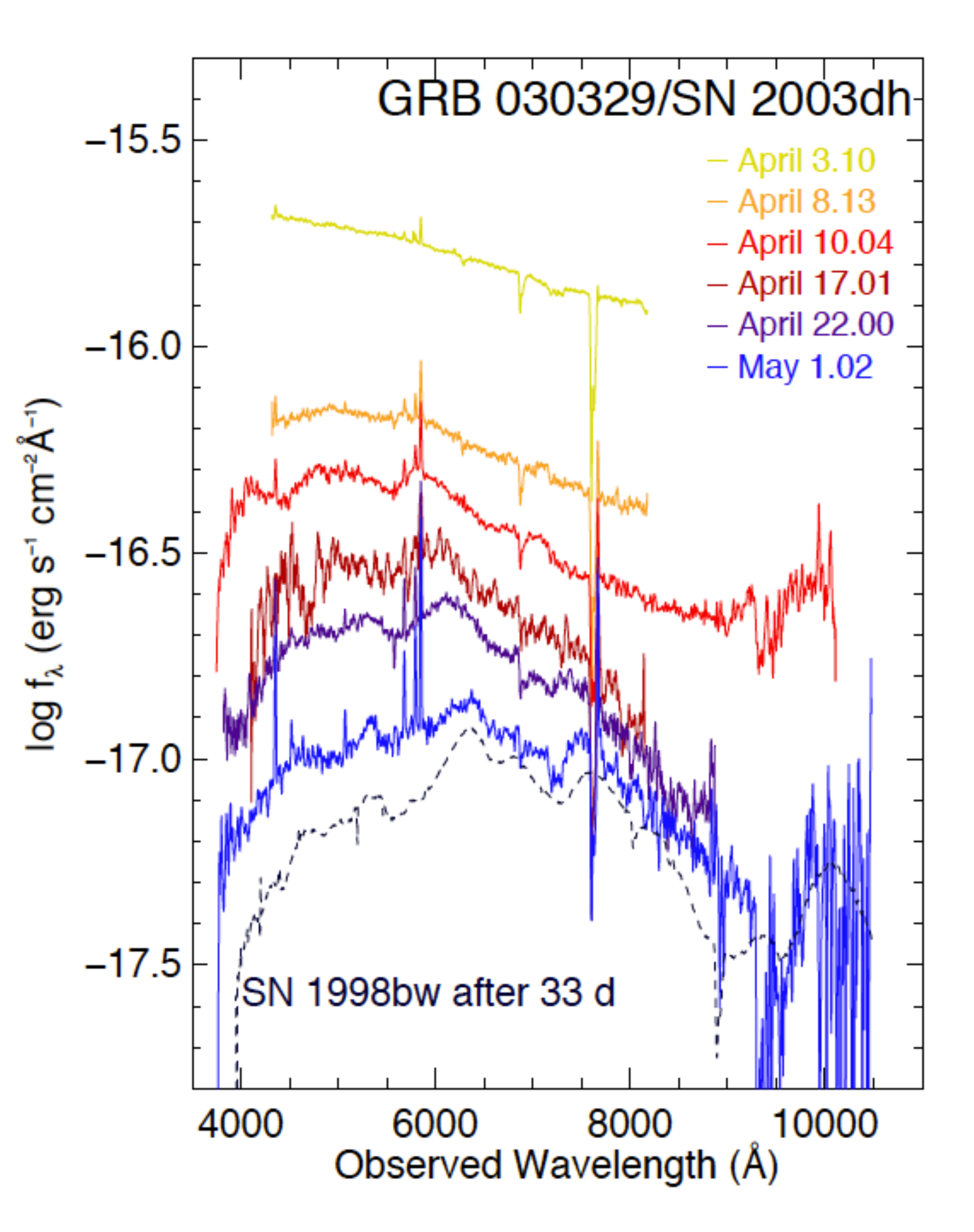}
\caption{Left: late SN bump is visible at $\sim 30 $ days after the GRB in the optical light curve of GRB~090618 in two filters, $R_{\rm c}$ (red) and $i$ (blue). The solid line is the best fit model with afterglow and SN1998bw-type SN, dimmed by 0.75 and 0.5 mag in $R_{\rm c}$ and $i$, respectively \cite{cano2}. Right: flux-calibrated spectra of the optical flux of GRB~030329. The spectrum evolves from a featureless power-law spectrum to a SN-like spectrum dominated by broad features. Superimposed on the spectrum are several strong emission lines from the underlying host galaxy \cite{hjorth}.}
\label{grb-sn}
\end{center}
\end{figure}

Among the first models suggested to explain GRBs were supernova (SN) explosions \cite{colgate2}. Indeed, total energy emitted in a GRB is roughly of the same order of magnitude as the energy liberated during a supernova explosion, but there are important differences between these types of event: in a supernova, the energy is emitted in a few months, but in a GRB in a matter of seconds; supernova outflow is non-relativistic and radiation is thermalized, while GRB outflow is relativistic, and emission is non-thermal. Therefore, it came as a great surprise when supernova SN1998bw was found in the error-box of GRB~980425 and indicated a possible connection between the two types of explosion. In the following five years, photometric evidence for GRB-SN connection emerged in the late light curves of a handful of optical afterglows \cite{hjorthbloom}: at typically $\sim 10$ days after the GRB itself, a bump appeared in the otherwise power-law decay of the optical afterglow (Fig. \ref{grb-sn} left). In 2003, the strongest evidence for the GRB-SN connection was observed: several days after GRB~030329, the spectrum of the optical afterglow showed emerging change from a featureless power-law spectrum, characteristic of GRB afterglows, to include more and more supernova features. By subtracting the afterglow spectrum, the SN spectrum was isolated, and it closely followed the broad-lined Type Ic SN spectrum of SN1998bw (Fig. \ref{grb-sn} right). It is now generally accepted that (at least some) long GRBs are connected to core collapse supernovae (i.e. supernovae triggered when cores of evolved massive stars collapse).

\subsection{GCN and rapid follow-up observations}
With the discovery of GRBs' afterglows, the need for the quick dissemination of the information about a new GRB became evident. For this purpose, the Gamma-ray burst Coordinates Network (GCN) was set up in 1997 \cite{barthelmy1}. Through the GCN, astronomers exchange information about follow-up observations, and most importantly, all interested observatories can receive socket-fed GRB alerts from satellites and slew their instruments to a GRB location in the sky without human intervention. Images taken on the ground are compared with existing star catalogues to determine whether there is a new source, i.e. an afterglow detected. Some telescopes have the whole process of imaging, afterglow detection and triggering the best sequence of follow-up observations completely automated (e.g. LT-TRAP \cite{guidorzi}). It is crucial to start the observations as soon as possible and catch the afterglows before they fade, to study the phenomena in the vicinity of the explosion, and, if possible, observe longer wavelength light curves simultaneously with high energy emission. A simultaneous multi-wavelength picture is needed to test different aspects of GRBs' nature and compare them to theoretical models. The GCN network is the backbone of multi-wavelength GRBs' observations, and enables the follow-up observations to start immediately (sometimes in less than one minute) after the GRB trigger. This became especially fruitful in the era of {\em Swift} satellite.

\subsection{Swift era }
\begin{figure}
\begin{center}
\includegraphics[angle=0, scale=0.50]{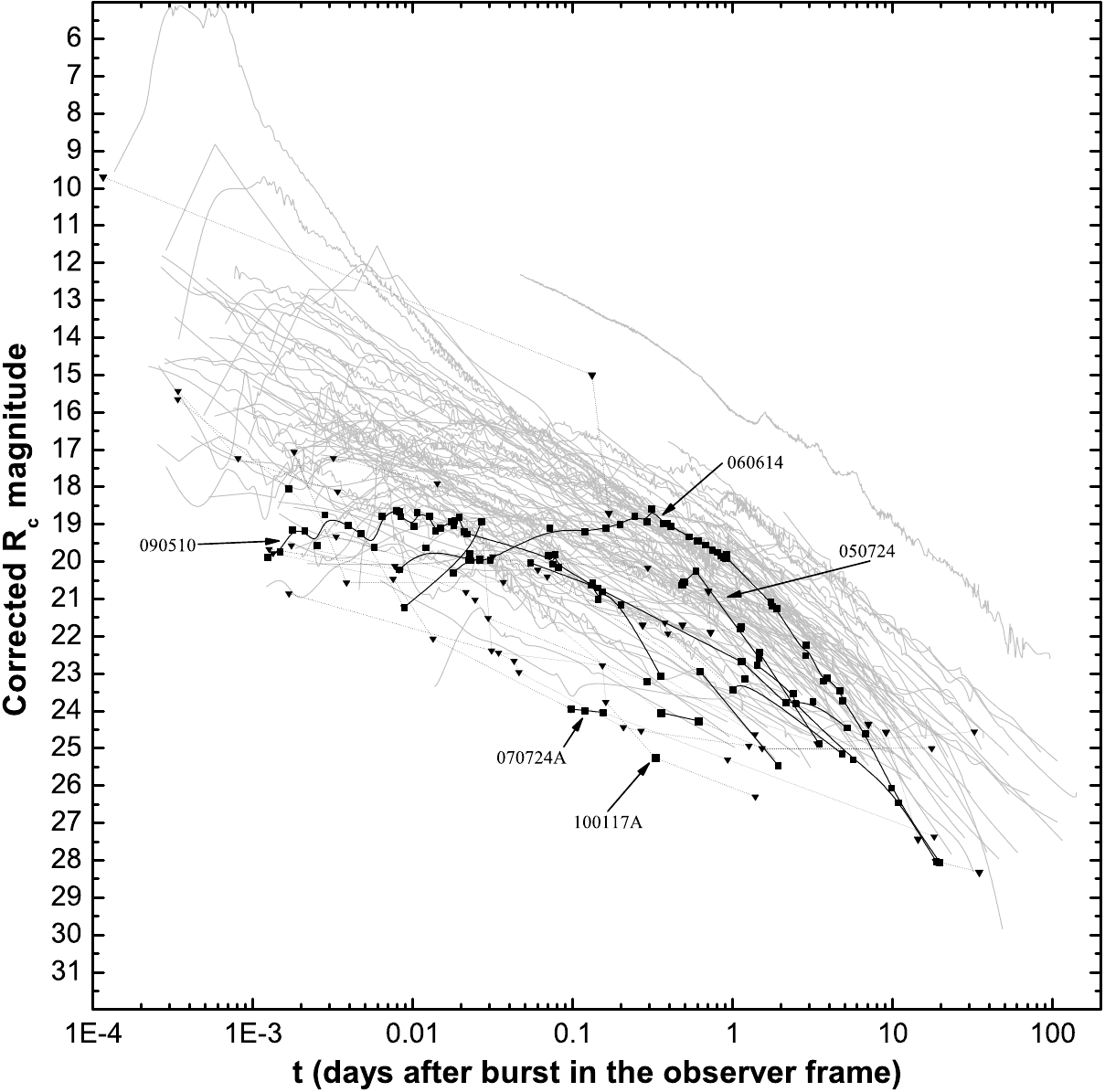}
\includegraphics[angle=0, scale=0.30]{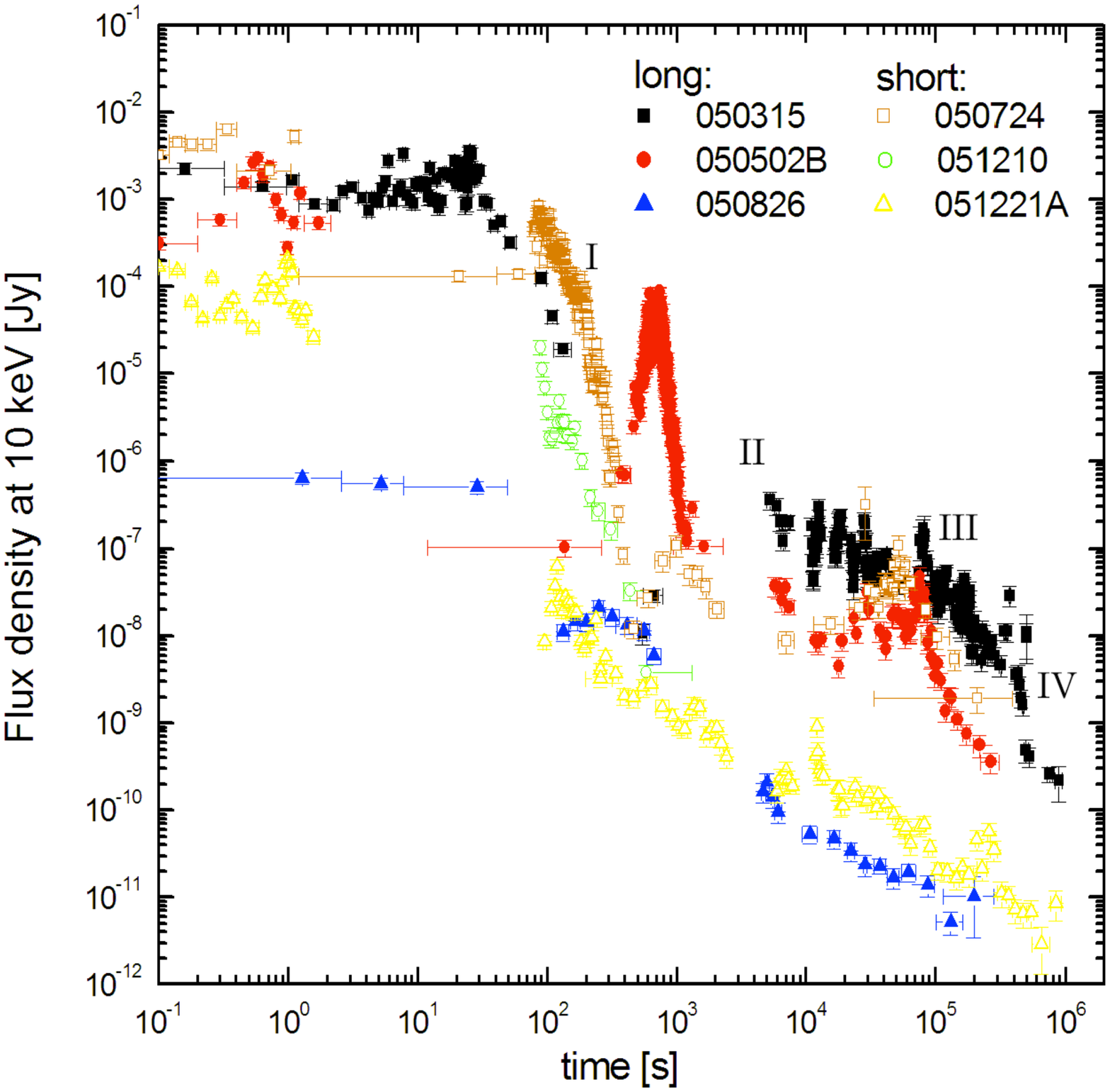}
\caption{Left: collection of optical afterglows of GRBs. Thin gray lines are long GRB afterglows; black lines with data points are upper limits (thin straight dashed lines, downward pointing triangles) or detections (splines, squares) of short GRB afterglows \cite{kann2}. Right: representative examples of X-ray afterglows of long and short GRBs with steep-to-shallow transitions (GRB 050315, 050724), large X-ray flares (GRB 050502B, 050724), fast declining (GRB 051210) and gradually declining (GRB 051221A, GRB 050826; flux scale divided by 100 for clarity) afterglows.}
\label{kann}
\end{center}
\end{figure}

{\em Swift} was launched in late 2004 and is specifically for GRBs' exploration built spacecraft. It has three main instruments: when the gamma ray detecting wide-field Burst Alert Telescope detects a GRB, it sends a trigger to the ground and GCN receivers. The X-ray Telescope is immediately slewed to the GRB's position and pinpoints the X-ray afterglow's position to a typical accuracy of 5 arcsec. This position is usually sent to receivers within one minute, and is followed a few minutes later by the image taken with the {\em Swift}'s third instrument, the Ultra Violet Optical Telescope.  High-quality observations of hundreds of GRBs, with better temporal and multi-band coverage of afterglows enabled by {\em Swift} has revealed an unforeseen richness in burst properties \cite{gehrels}. It has led to the detection of short GRBs' optical afterglows, shown the diversity in optical afterglows (Fig. \ref{kann} left) and the potential of GRBs to be used as probes of their environment out to high redshifts (see section \ref{environment} and \ref{cosmo-probes}).

One of the breakthrough discoveries by {\em Swift} is the X-ray afterglow behavior in the first few hours after a GRB, which were missed by {\em Beppo-SAX}. {\em Swift} revealed several striking features: (i) many early X-ray light curves show a canonical behavior with three distinct power-law segments (marked I - III in Fig. \ref{kann} right), in some cases also a jet break at later times (IV); (ii) in about half of the GRBs, bright flares in the X-ray light curves are observed long after the end of the prompt phase ($10^2$ s - $10^4$ s). In some extreme cases, these flares have integrated energy similar to, or exceeding, the initial burst of gamma rays, and severely challenge current theoretical models. 

\subsection{Fermi era}
While the {\em Swift} satellite produced many important new findings, particularly in afterglow physics and the study of GRBs' environment, {\em Fermi} satellite (launched in 2008), has focused attention on GRBs' prompt emission phase. {\em Fermi} carries two types of gamma detector (GBM and LAT), which cover a much broader energy range (8 keV-300 GeV) than {\em Swift} (15-150 keV), and give us a much wider picture of prompt gamma emission. {\em Fermi} results show that most GRBs have familiar broken power-law spectra, as already observed by BATSE. However, several GRBs detected in very high energy regimes show evidence of a separate extra high-energy component which can not be fitted with a single broken power law function. In several cases, {\em Fermi} also detected a significant thermal component in addition to the dominant non-thermal components. Also very interesting are the observations which show that in most (but not all) GRBs, high energy photons arrive with a delay of a few seconds compared to lower energy photons, and are frequently observed until a few seconds later than low energy photons. In short, {\em Fermi} revealed the rich phenomenology of the prompt emission phase, which does not fit any current theoretical model. 

\section{Progenitor Models}
After reviewing the main observational discoveries, we now discuss their interpretation: which objects produce GRBs and how we can understand in terms of the physics what causes and drives them. It is generally considered that the study of possible progenitors and the description of a GRB explosion itself can be de-coupled: after a huge amount of energy is released in a brief time in a small volume, the explosion that follows is not very dependent on the details of its progenitor and central engine. So, in the next section, we first consider the progenitor models, about which we are reasonably certain, and then discuss GRB explosion models, about which we are less certain.  Inevitably, most of more than a hundred models proposed early on to explain GRBs' origin were abandoned as new properties of these events became known, but a few have survived. The most accepted models for GRBs' progenitors are now the collapsar model for long GRBs and the compact binary merger for short GRBs. 

\subsection{Collapsar model}
Based on the observational evidence connecting GRBs to core collapse supernovae, it may be said that (at least some) long GRBs are caused by a collapse of massive stars. 

The starting point for the collapsar model is a massive, rapidly rotating Wolf-Rayet star\footnote[4]{Wolf-Rayet stars are a normal stage in the evolution of very massive stars. They are very hot, evolved, massive stars (over 20 $M_\odot$ initially), which are losing mass rapidly, typically $10^{-5} M_\odot $/year, by means of a very strong stellar wind.}, which has lost its hydrogen envelope, has a core with about $10 M_\odot$ and is the size of the Sun \cite{woosley}. When nuclear fusion reactions in the core stop, the core becomes unstable, and it collapses to a black hole, with a few $M_\odot$, surrounded by a massive accretion disc. In the equatorial plane (perpendicular to the star's rotation axis), the accretion of the surrounding envelope of stellar matter is halted, since the material has too large angular momentum to fall directly into the black hole. Along the rotation axis, however, it can undergo almost free-fall; this rarifies the region around the rotation axis and makes a 'low density funnel' in the envelope. If enough energy is injected into this region, it is able to push material along the rotation axis for as long it takes to cross the star, typically a few $\sim 10$ s, and the outflow eventually breaks through the star's surface. Numerical simulations show that in this case the outflow is collimated by the pressure from the stellar mantle, and gains high Lorentz factors as it breaks through the surface, forming a collimated, relativistic outflow or a jet. It should be noted that the collapsar model does not specify {\em how} the jet is launched, but merely assumes that it is in one way or another.\footnote[5]{Jets are encountered on many different scales throughout astrophysics, from young stars to neutron stars, gamma ray bursts and active galactic nuclei. The exact mechanism of their launch is not well known, but it seems plausible that it is related to the accretion disc and magnetic field.}

The energy can come from neutrinos produced in the accretion disc and the annihilation $\nu + {\bar {\nu}}\rightarrow  e^+ + e^-$. Neutrino energy comes from accretion, and in general accretion powered luminosity is $L_{\rm acc} = \zeta \dot{M}c^2$, where $\zeta$ is the efficiency factor and $\dot{M}$ the mass accretion rate. To achieve the observed GRBs' luminosities, $\dot{M}$ should be $\sim 0.1 M_\odot /{\rm s} - 1 M_\odot /{\rm s}$ for reasonable $\zeta$. Other possibilities include extraction of the rotational energy of the black hole through strong magnetic fields in the accretion disc, which would require a rapidly spinning black hole and high accretion rate, creating strong magnetic fields ($B\geq 10^{11}\, $T).

\begin{figure}
\begin{center}
\includegraphics[angle=0, scale=0.4]{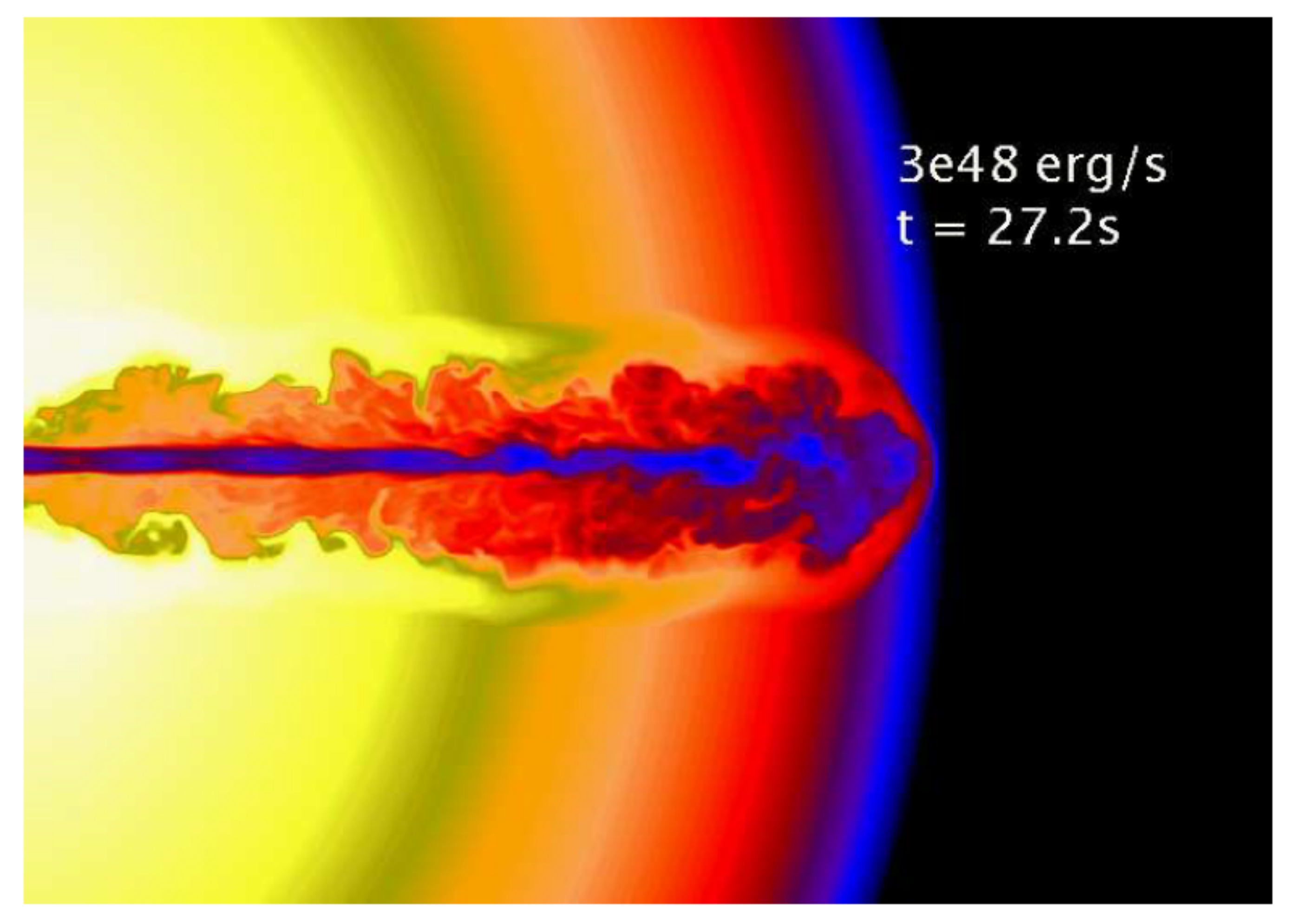}
\caption{Three dimensional modeling of a relativistic jet with energy $3\cdot 10^{41}$ J introduced at $10^8 $ m from the centre of a $15 M_\odot$ Wolf-Rayet star with radius $8 \cdot 10^{8} $ m. Plotted is the logarithm of the density as the jet nears the surface \cite{woosley}.}
\label{jet-sim}
\end{center}
\end{figure}

It is expected that only a small fraction of core collapse supernovae produces long GRBs, with their rate around $10^{-7}$/galaxy/yr (compared to the much higher rate of supernovae Type Ic). On the other hand, although direct connection with SN was observed for only a few tens of long GRBs, it is generally assumed that {\em all} long GRBs are triggered by the death of massive stars. Massive stars have a very short lifetime of $\sim 10^7$ yr; they reach their ends while star formation in their vicinity is still active and before they can move far from their birth places. If long GRBs are produced by the collapsar model, then they should be found in regions of intensive star formation. Indeed, detailed observations of GRBs' host galaxies and GRBs' locations within them reveal that they lie in star forming regions of blue dwarf galaxies.

\subsection{Compact binary mergers}
Binary systems of two neutron stars and of a neutron star and a black hole are mutually referred to as compact binary systems.\footnote[6]{Both neutron stars and stellar mass black holes are remnants of massive stars, and are compact - very small considering their mass. Neutron stars have masses between 1.4 $M_\odot$ and 2 $M_\odot$ and are only about 10 km in size; their densities are therefore of the order of a few $10^{17} {\rm kg/m^3}$, i.e. comparable to the  density of an atomic nucleus. Rapidly rotating young neutron stars with strong magnetic fields emit a beam of electromagnetic radiation, and are called pulsars. Stellar remnants with masses larger than $\sim 2 M_\odot$ are thought to collapse to a black hole.} It was suggested that they were GRB engines early on \cite{eichler} and they are now the most popular model for the central engines of short GRBs.

\begin{figure}
\begin{center}
\includegraphics[angle=0, scale=0.3]{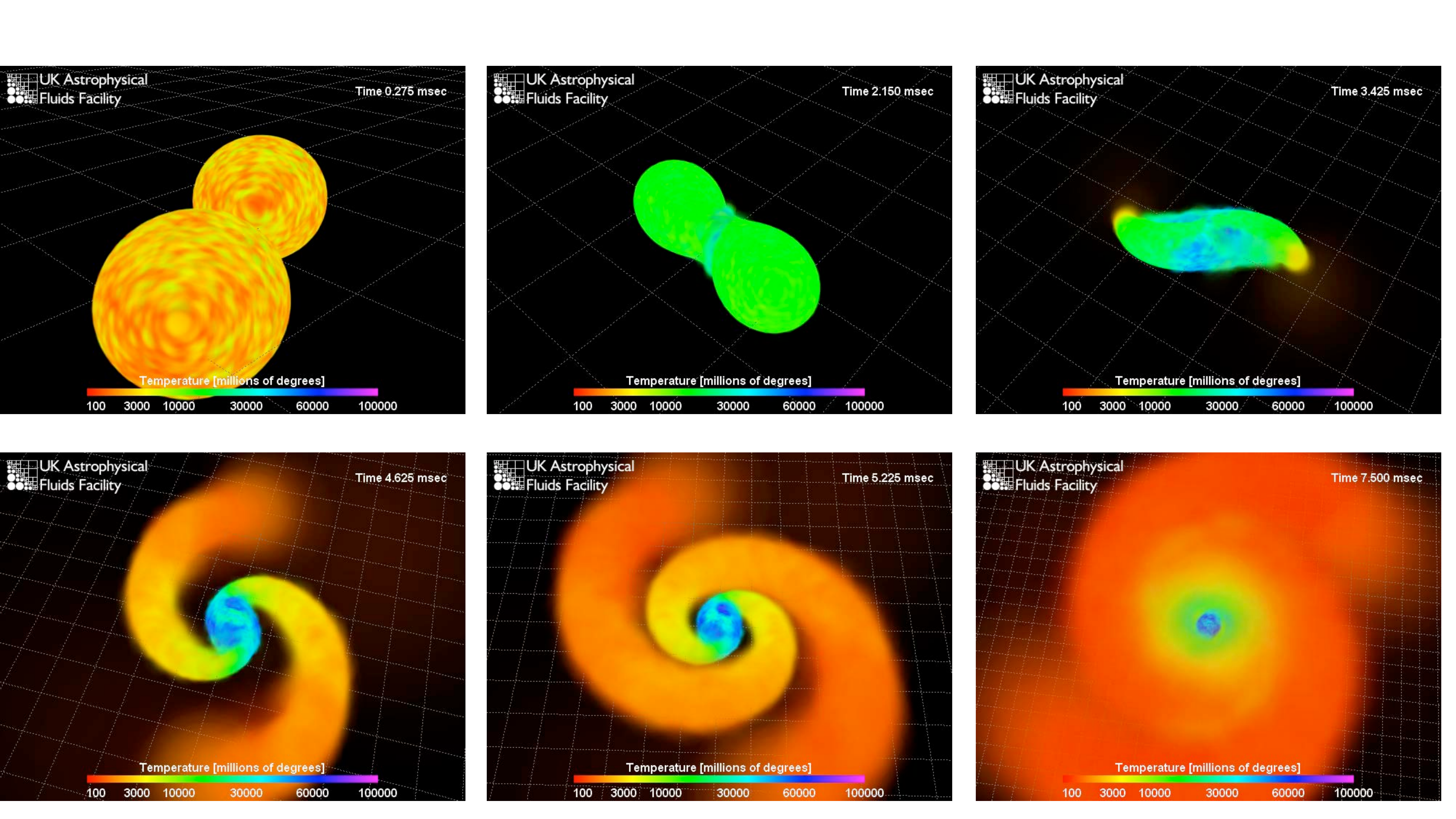}
\caption{Snapshots of simulation of two neutron starsÕ merger (each neutron star has  $1.4 M_\odot$ and $\approx 30$ km diameter). Initially, they are less than 10 km apart, and moving at around $v=0.2 c$. As the two stars spiral together, they become deformed, and finally touch. When they merge, the matter reaches $T\sim 10^{11}$ K. A few percent of the matter is ejected in the form of spiral arms, which cool rapidly. The whole merger process takes only a few ms. The grid in the images has a spacing of 30 km intervals. Credit: simulation by Stephan Rosswog, visualization by Richard West, http://www.ukaff.ac.uk/movies/nsmerger/}
\label{merger}
\end{center}
\end{figure}

The merger process of two neutron stars begins with the slow inspiral phase, which can take from only $\sim 10^6$ yr to $\sim 10^9$ yr. During the inspiral, the system emits gravitational waves, and the orbital period and separation of stars decrease. Support for this picture comes from the fact that such systems are indeed observed: the orbital change of the binary pulsar PSR B1913+16 is in excellent agreement with general relativity predictions for the system, which is losing energy by the emission of gravitational waves.\footnote[7]{For this discovery, Russell A. Hulse and Joseph H. Taylor  received the Nobel Prize for Physics in 1993.} The last stages of the inspiral occur very rapidly, with the final 100 km taking less than a second. Once neutron stars approach within a few of their radii, tidal interaction distorts their shapes, immediately before they merge within $\sim $ms (Fig. \ref{merger}). Excess angular momentum is carried by two long spiral arms wrapped around the central object, which collapses to a black hole. In the final stage, the configuration is similar to the central region of a collapsar: a low-mass black hole surrounded by a massive accretion disc.

The merger of a neutron star and a black hole proceeds similarly, but is more complicated due to the transfer of mass from a neutron star to a black hole. 

Also in the merger model, there are two possible sources of energy: gravitational and rotational. The outflow is probably launched similarly as in the collapsar model: in the region above the poles of the newly formed black hole. It is also expected that the magnetic fields of (one or two) neutron stars are substantially amplified during the merger, and could play an important role in the launch of the outflow. 

According to this scenario, we can expect to find short GRBs both inside and outside galaxies and their star forming regions. If we take into account that the inspiral time for a compact binary can be quite short, we may expect to find them still inside regions where their progenitor stars were formed and star formation is continuing. On the other hand, it is known that neutron stars can experience substantial kicks when they are formed during a supernova explosion. If this is combined with a long inspiral time, it is possible that, by the time they merge, they have traveled far outside star forming regions, and even outside their galaxies. Indeed, observations show that short GRBs are found in both types of environment: inside galaxies (in star forming regions of young galaxies and in old galaxies with low star formation), but also far outside galaxies. However, contrary to long GRBs, for short GRBs we do not have a 'smoking gun' (such as the supernova signature for long GRBs) to identify with certainty the nature of their progenitors.

\section{Theoretical models of GRBs and their afterglows}
What exactly happens in a GRB explosion? The physics involved in describing GRBs from 'first principles' is quite complicated and there are several uncertain intermediate steps relating the physics of the central engine to the properties of the relativistic jet and the gamma-ray and afterglow emission mechanisms. As already mentioned, it is generally considered that, whatever the progenitor, the evolution of the explosion after tremendous energy injection can be described independently of the details of the progenitor. This is supported by the non-transparency of the central region of the explosion and the so-called compactness problem.

\subsection{Compactness problem}\label{cp}
The compactness problem arises due to the combination of the large energy involved, short-time variability and observed non-thermal spectrum. The short time variability (on a timescale of $\Delta t$) implies, due to causality, small source size: $D \leq c\cdot \Delta t$. But such a compact source with large luminosity in gamma rays would be opaque to its own radiation, because photons would have high enough energies to create a large number of $e^-e^+$ pairs. Therefore, the end result should be a thermal spectrum, contrary to observations.

Let us make a very crude estimate of the optical depth in such an energetic and compact source: $\tau \sim n_e \sigma_T D$ (where $n_e$ is the number density of free electrons, and $\sigma_T$ is the Thomson cross-section). If we assume that the emission is isotropic\footnote[8]{Taking into account the collimation of the outflow does not change the conclusion.}, the involved energy is $E_{\rm iso}$. With typical observed photon energy ${\bar E_\gamma } \sim 1\, $MeV, photon number density at the source is about $n_\gamma \sim {{E_{\rm iso}}\over {{\bar E_\gamma }D^3}}$. Because not all photons will produce $e^-e^+$ pairs, we denote electron density as $n_e = f_e \cdot n_\gamma $, and we can write the optical depth as:
\begin{equation}
\tau \sim {{10^{13} f_{\rm e} \Biggl({ {E_{\rm iso}}\over {10^{42}\,  {\rm J}}}\Biggr) \Biggl({{1\, {\rm MeV}}\over {\bar{E_\gamma}}}\Biggr) {\Biggl({ {0.01\, {\rm s}}\over {\Delta t}}\Biggr)}}}^2 \label{comp}
\end{equation}
It is evident that for any reasonable set of parameters, $\tau $ is much larger than 1, i.e. the source is optically thick and the resulting radiation should be thermal. However, there is a flaw in the above estimate if the motion of the outflow is relativistic. If there is variability in the relativistic outflow on a timescale $\Delta t_{\rm em}$, the observer, to whom the outflow is approaching at a relativistic speed with Lorentz factor $\Gamma$, measures the variability timescale to be $\Delta t_{\rm obs}={{\Delta t_{\rm em}}\over {2\Gamma ^2}}$ (this effect is similar to super-luminal motion; for details of the derivation see \cite{rosswog} or \cite{meszaros}). Therefore, if the outflow is relativistic, the source size can be larger than $c\Delta t_{\rm obs}$ by a factor of the order of $\Gamma^2$. The second effect is the shift of photon energies: the observed photon energy is $E_{\rm \gamma, obs}= \Gamma E_{\rm \gamma , em}$. In fact, only a small fraction of photons is energetic enough in the outflow rest frame to produce $e^-e^+$ pairs. Their fraction depends on the shape of the spectrum, and detailed calculations show that our crude estimate needs to be corrected for a factor of $\Gamma ^{-6}$ \cite{lithwick}. To get $\tau \sim 1$ in eq. (\ref{comp}), we need as a lower limit $\Gamma \sim 100$: in order for observers to receive high-energy, non-thermal spectra of gamma rays, relativistic motion with at least $\Gamma \sim 100$ is required.

\subsection{General picture}
The basic picture of a GRBs model is as follows: a stellar mass object undergoes a catastrophic event, which releases a large amount of energy in a small region, about $10 - 100$ km in size. The released energy can be in the form of radiation, thermal energy and/or electromagnetic energy, and it drives the subsequent acceleration to highly relativistic velocities. Since this flow is initially optically thick, it undergoes adiabatic expansion and cools. At large distances, $> 10^{11}$ m $\sim 1$.a.u. from the central engine, it becomes optically thin, and only then can gamma ray photons escape. It is this radiation, produced at large distances from the central engine, which we see as a GRB, and not the initial release of energy. Emitted gamma photons carry away only a fraction of the energy of the outflow. The rest is carried on, to distances of $\sim 10^{14} - 10^{16}$ m $\sim 0.01\,  {\rm lyr}- 1\, {\rm lyr}$, where the outflow collides with the surrounding medium and produces electromagnetic radiation at lower frequencies, called the afterglow.

 \begin{figure}
\begin{center}
\includegraphics[angle=0, scale=0.09]{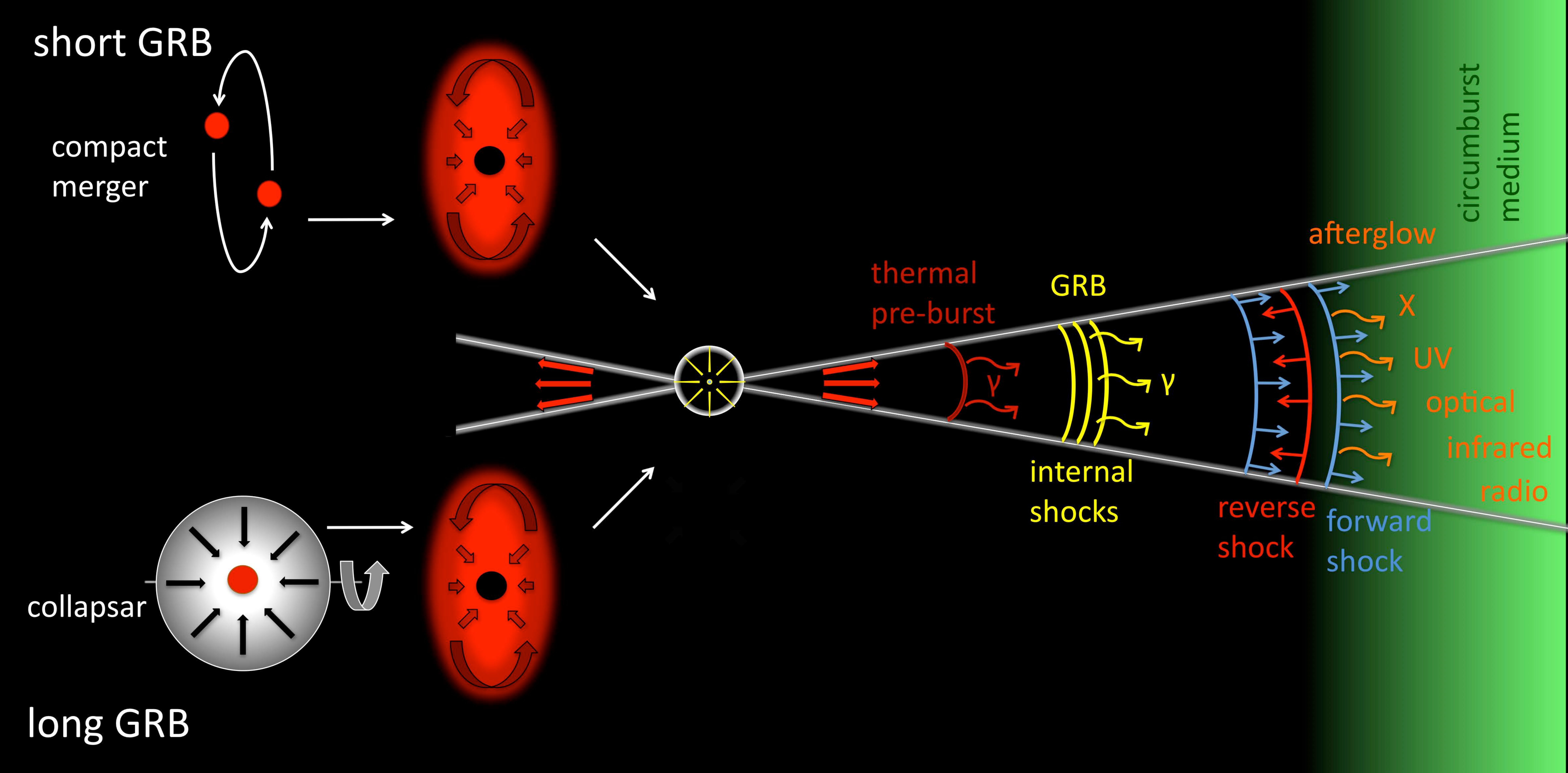}
\caption{Progenitor models for short and long GRBs (left). Production sites of $\gamma$ and afterglow emission in the fireball model (right).}
\label{fbm}
\end{center}
\end{figure}

While this general picture is largely agreed on, there are many important steps in it which remain open to question, the most crucial being the composition of the outflow. The outflow is composed of three components: matter, magnetic fields, and photons. Photons decouple from ejecta when it becomes transparent, and the jet carries on matter and magnetic flux. More photons are generated in regions where kinetic or magnetic energy is dissipated (in shocks or magnetic reconnection regions) and escape without further coupling. This leaves us with matter and magnetic field - and the question: which of them is the dominant ingredient of the outflow? Let us denote the distribution of the energy between magnetic field and matter by the ratio between the magnetic and matter flux, i.e. the magnetization factor $\sigma = B^2/4\pi \Gamma\rho c^2$ (where $B$ is the magnetic field and $\rho$ is matter density). Currently, there are two main types of model, which differ in the role they ascribe to the magnetic field: \begin{itemize} 
\item{The traditional {\em fireball model} \cite{sari, piran} is matter dominated, $\sigma \ll 1$, and magnetic field does not play a kinematically dominant role in it. During the expansion, the fireball transforms most of its thermal energy into the kinetic energy of its baryons, which become highly relativistic. If the outflow is not completely homogeneous, the expanding plasma is structured as several shells with slightly different Lorentz factors, with the faster shells catching up with the slower ones and generating {\em internal} shocks. In these, which are believed to be collision-less plasma shocks with amplified magnetic field, electrons emit via a synchrotron process, in some cases perhaps also inverse Compton processes - this is the prompt GRB emission. As the fireball expands, it eventually impacts on surrounding material and dissipates kinetic energy in so-called {\em external} shocks, which are created between the outflow and external material. The electrons in the external shocks are accelerated and produce synchrotron emission in locally generated magnetic fields - the afterglow. The spectral energy distribution of this synchrotron emission has several power law segments and, as the flow slows down, the peak of the emission shifts to lower energies, from X-rays through ultraviolet, optical, infrared to radio wavelengths. The fireball model also predicts that the afterglows will fade with time according to a power law - in general, therefore: $F_\nu (t) \propto t^{-\alpha } \nu^{-\beta}$  - and depend on several parameters, e.g. electron energy distribution, surrounding medium profile, energy refreshment etc. 

The external shock produced is typically forward shock. But if the ejecta plasma is weakly magnetized, the collision of the fireball with the surrounding medium can also give rise to a reverse shock, traveling backwards through the expanding fireball, and producing a bright and rapidly fading optical emission called optical flash. 
}
\item{In {\em electromagnetic models} $\sigma >1$, plasma is strongly magnetized, and the ejecta carries a globally ordered magnetic field, which is kinematically important. There are several models: one model \cite{lyutikov1} assumes that the energy that powers a GRB comes from the rotational kinetic energy of the central source, and is extracted through a magnetic field, which can be generated by a local dynamo mechanism. An important difference with regard to the fireball model is that the energy is dissipated directly into emitting particles through current-driven instabilities, not through shocks. The magnetohydrodynamical model is similar \cite{drenkhahn}, although in this case the magnetic field energy is first converted into bulk motion and then dissipated through internal shocks, as in the fireball model. Recently, the internal-collision-induced magnetic reconnection and turbulence model was proposed, which, as the name suggests, includes quite complicated physics \cite{zhang2011}.}
\end{itemize}

\section{Open issues in GRBs physics}
In spite of great theoretical efforts and successes, many fundamental questions remain about the physics of GRBs. It is not within the scope of this paper to go into detail about them; for an excellent review and discussion, we refer the reader to the paper by Zhang 2011 \cite{zhang11}. Here we only briefly sketch them.
\begin{itemize}
\item{The {\bf classification} of GRBs into long and short, as based on their duration in gamma rays, $T_{90}$, has several weaknesses. $T_{90}$ is an observer's frame quantity (not an intrinsic, GRB frame quantity) and depends on the energy band and sensitivity of the detector. In addition, several GRBs have been detected which had some of the characteristics of short type, and some of the long type. Therefore, classification which also takes into account afterglow and host galaxy properties seems more justified, and some suggest more 'physical categories' connected to the nature of the progenitor: 'Type I GRBs' (i.e. due to compact mergers) and 'Type II GRBs' (i.e. due to the collapse of massive stars).
}
\item{{\bf Progenitors} of GRBs are not directly observed. To identify them indirectly, theoretical and observational approaches are used: observational data are used to narrow down allowed progenitor types, and detailed theoretical studies investigate whether GRBs can be produced and under which circumstances. 

Regarding short GRBs and compact binary mergers, we have no direct observational link between them, only indirect, based on their environment, types of host galaxy and GRBs locations in them. We know that neutron star-neutron star binaries exist and that their orbits shrink due to gravitational radiation, so they are doomed to merge eventually. However, there are several observational facts which are difficult to explain theoretically. One of these is the extremely narrow beaming required by some observations, while numerical modeling of compact mergers consistently produces wide jets. Definite proof of this scenario would be the detection of specific 'compact merger' gravitational wave signals with one of the gravitational wave detectors (e.g., LIGO).

The connection between long GRBs and the death of massive stars is more direct. To date, about 20 cases of significant SN bumps in the light curves and 5 firm cases of spectroscopic association of GRBs with core-collapse supernovae have been observed. But as it becomes difficult to detect supernova features with increasing distance, the current sample of GRB-SN connection events is limited to relatively nearby events ($z_{\rm max}=1.058$), all being long GRBs. There are, however, also a few cases of long GRBs occurring close enough for SN to be observed, but it was not in spite of deep observations. On the other hand, various SN surveys are now detecting a growing number of broad line SN IcÕs, which are not observed to have accompanied GRB. Although some of these events could be due to an 'inconvenient' viewing angle (if our line of sight does not intersect the collimated GRB jet, we may not detect a GRB, but we may detect a SN or even an 'off-axis' or 'orphan' afterglow \cite{japelj}), the statistics and stellar evolution models indicate that the majority of SNs do not produce GRBs. At the moment, we do not understand what distinguishes GRB/SN, GRB/SN-less and SN/GRB-less progenitors.  One obvious culprit could be one of the most fundamental properties of a star, its mass, rotation or metallicity\footnote[9]{In astronomy, metallicity measures the content of all chemical elements other than hydrogen and helium.} \cite{modjaz}. Many theoretical models favor rapidly rotating massive stars with low metallicity as likely GRB progenitors.}

\item{The {\bf central engine} in both the collapsar and compact merger models is usually considered to be a newborn (possibly rapidly rotating) black hole surrounded by a massive accretion disc. In some cases, it can also be a rapidly rotating young neutron star endowed with a strong magnetic field, i.e. a magnetar, similar to those  magnetars, which are found in young supernova remnants. In order to power a GRB, the magnetar, which may also be surrounded by an accretion disc, must have a strong surface magnetic field, $B \gtrsim 10^{11}$ T, and a spin period $P \lesssim 1$ ms. The main power of such a millisecond magnetar would be its spindown power. A more exotic possibility is a quark star, i.e. a compact star composed of quark matter. In this scenario, the energy reservoir includes, in addition to the accretion power and rotational energy, the energy due to phase transitions (e.g. transition from neutron matter to matter made of degenerate u, d quarks or from u, d quark matter to u, d, s quark matter). Unlike black holes and magnetars, the existence of quark stars has not been observationally confirmed. But as all three types of central engine are argued to satisfy most observational constraints, it is not straightforward to identify the correct one.}

\item{{\bf Energy dissipation and particle acceleration mechanism}:
it is generally agreed, that the energy of a GRB comes from the gravitational and/or rotational energy of the central engine. However, how a fraction of this energy accelerates particles, which then subsequently radiate, is still a matter of debate. Whether the acceleration of particles is due to shocks, magnetized shocks or magnetic dissipation is subject to further investigations.
}

\item{{\bf Radiation mechanism}:
 the prompt emission of GRBs is evidently non-thermal and requires a population of particles (likely electrons) with a power-law distribution. Three radiation mechanisms have been considered for these particles: synchrotron radiation, synchrotron self-Compton radiation, and Compton up-scattering of thermal seed photons (for instance, from the photosphere). All suffer from several weaknesses, and it is difficult to identify the correct radiation mechanism with currently available data. The origin of afterglow emission is relatively better understood. Broad band afterglow spectrum can usually be modeled by synchrotron and synchrotron self-Compton radiation from electrons in the external shock.
}
\item{The {\bf origin of the high energy emission}: based on recent {\em Fermi} results, prompt GRBs' emission can have three components: usual broken power-law component, thermal component and an extra, high energy power law component. Interpretation is sometimes difficult because of contradictory results. The high energy emission detected in a few GRBs is highly debated: to explain GeV emission, some propose a second episode of energy dissipation above the photosphere, while others propose a different origin of the high energy component, i.e. that it comes from hadrons, not electrons. 
}
\item{{\bf Long term central engine activity}:
the results of {\em Swift} observations of X-ray afterglows (Fig. \ref{kann} right) present serious challenges for GRB models. While phases I and III can be interpreted in the fireball model as off-axis emission from regions at $\theta > 1/\Gamma$ and normal afterglow, the shallow phase II and X-ray flares are more difficult to explain. Most of the proposed models include long lasting (up to $\sim 10^4$ s after the prompt emission phase is over) central engine activity, which should explain both erratic X-ray flares and plateaus, which require a smooth energizing component. Mechanisms to account for the latter include fall-back accretion on a black hole or tapping the rotational energy of a millisecond magnetar. The mechanisms proposed for erratic X-ray flares include the fragmentation of the collapsing star, fragmentation of the accretion disc, intermittent accretion behavior caused by a time variable magnetic barrier, magnetic bubbles, helium synthesis in the post-merger debris, and quakes in solid quark stars. Theoreticians in the GRB field certainly are creative.
}
\item{{\bf Afterglow physics}:\label{afph}
in pre-{\em Swift} era, the fireball model was very successful in describing smooth power law features of afterglows. This substantially changed in the {\em Swift} era, with early X-ray and simultaneous longer wavelength observations. As we have just seen, the behavior of X-ray afterglows presents a big challenge due to the great energies involved, but also detailed observational coverage of optical and infrared afterglows reveals more complex behavior \cite{kann2, kann1}. In addition to general power-law decay, many afterglows show features like the temporal luminosity enhancements interpreted as density bumps, and one or more peaks interpreted as due to refreshed shocks or late time energy injections. There are bursts, like GRB~080319B, with complex multi-wavelength behavior, for which a two-component jet (one narrow and very energetic, the other broader and slower) has been proposed. 

Simultaneous multi-wavelength observations have also put in doubt the interpretation of the late time steepening of the afterglow light curve as due to a jet break: some bursts have chromatic breaks (e. g. a break in optical, but not in X-rays or vice versa), or no break up to the very late time after the GRB, implying that the opening angles of jets are quite large. To resolve this issue, which puts severe constraints on theoretical models, the energy involved and number of GRBs, more late time observations are needed.

On the other hand, observations immediately after the GRB are crucial to studying the emission of the reverse shock, which propagates backwards through the expanding shell. Since the very bright optical flash of GRB~990123, it was expected that in the {\em Swift} era bright optical flashes would be prevalent in early-time afterglow emission, but rapid response observations show this not to be the case \cite{gomboc2009}. One possible explanation for the paucity of bright optical flashes is that the reverse shock emission peaks at longer wavelengths, perhaps as long as radio wavelengths \cite{melandri}, while the second explanation is magnetization of the outflow \cite{gomboc}. The reverse shock emission produced by magnetized outflows should be polarized, thus polarization measurements at very early times are of key importance for distinguishing between matter-dominated and magnetized outflow models. The feasibility of early optical polarimetry was demonstrated with the RINGO polarimeter on the robotic Liverpool Telescope \cite{mundell}, and in at least one case indicates the presence of a large-scale ordered magnetic field \cite{steele}.}

\item{{\bf Ejecta composition}:
 interest in magnetized outflows as a way to understand the dynamics and composition of GRBs has also gained momentum from recent results by {\em Fermi} and {\em Swift}. Magnetization is thus a fundamental issue in GRB physics, with implications in all areas of their understanding. Early polarimetry in the gamma, X-ray, and optical regime is of key importance in helping to resolve this issue, since it is an independent probe of the physical conditions of GRBs. The issue of magnetization is also linked to the question of whether GRBs are dominant sources of ultra-high energy cosmic rays and high-energy neutrinos. In the case of magnetized flux, the strength of these signals in GRBs would be lower by $(1+\sigma)^{-1}$, implying that GRBs are probably not the main contributors to high-energy cosmic rays and high-energy neutrino background.}
\end{itemize}

\section{Host galaxies and the environment of GRBs}\label{environment}
The study of the surrounding medium and GRBs' host galaxies can help us to better understand in what kind of environment GRBs explode and what their progenitors are. Using multi-wavelength afterglow observations, it is possible to study extinction in the line of sight. Usually, it is found that extinction profiles of Small or Large Magellanic Cloud fit the observations better than the Milky Way profile. Direct observations of host galaxies reveal even more about the birthplaces of GRBs. The host galaxies of long GRBs are mostly faint, blue, irregular, low-mass star forming galaxies with low metallicity. On the other hand, the host galaxies of short GRBs have larger luminosities and metallicities. Some short GRBs have been found in old and massive galaxies without recent star formation; however, the majority of short GRBs seem to explode in star forming galaxies. A detailed comparison between the hosts of short and long GRBs reveals systematic differences, and statistical tests show that they are not drawn from the same galaxy population \cite{berger}. 

\begin{figure}
\begin{center}
\includegraphics[angle=0, scale=0.4]{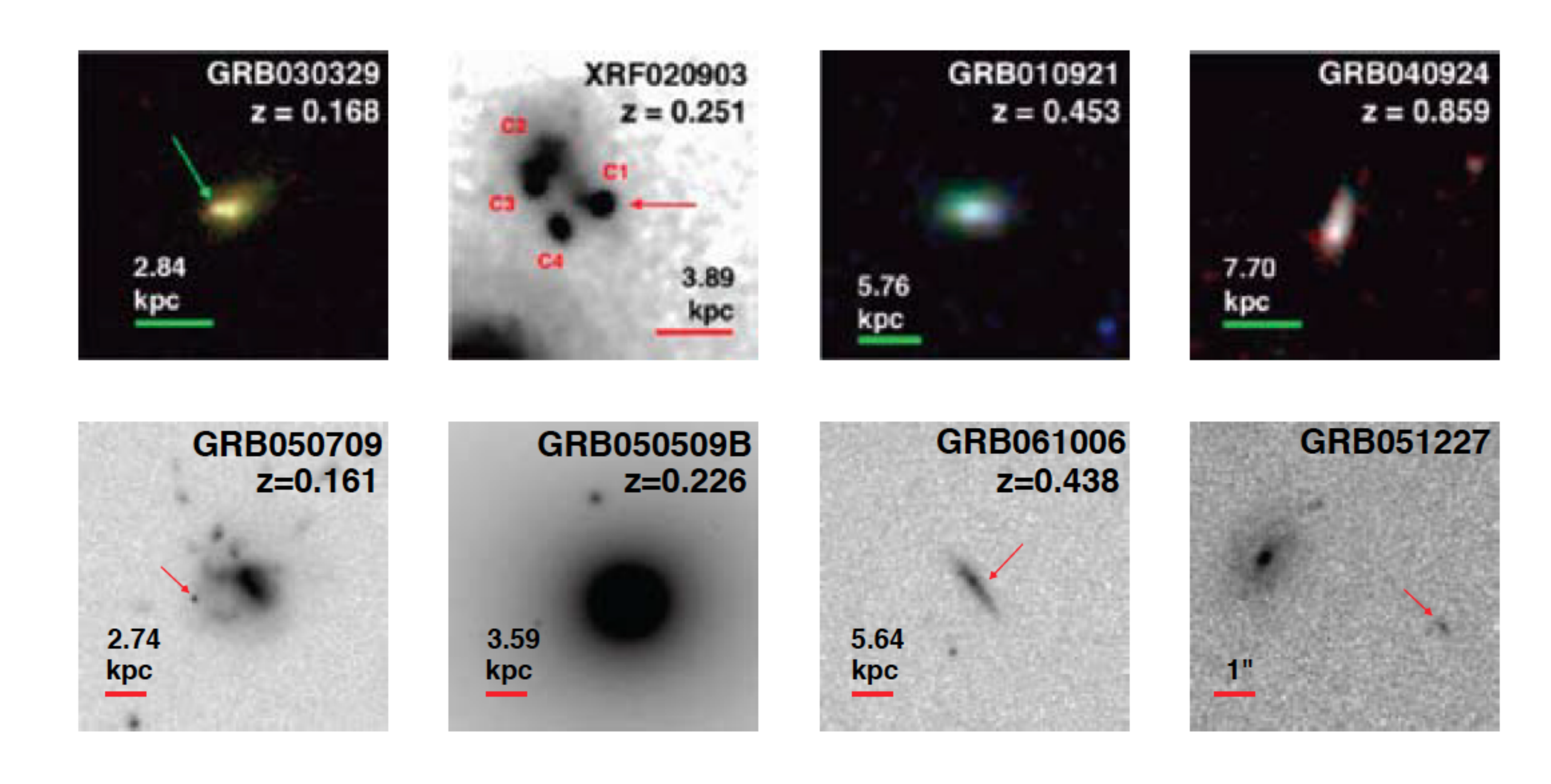}
\caption{A selection of the host galaxies of long (top row) and short (bottom row) GRBs, as imaged by the Hubble Space Telescope. Pairs of long and short burst host galaxies with comparable
redshifts were chosen. The physical length scale for a 1" angular distance is indicated in each panel (except for GRB~051227); arrows point to the location of the burst where this is known to pixel precision. Long GRBs' host images are from \cite{wainwright}; short GRBs' host images from \cite{fox} and \cite{gehrels}.}
\label{galaxies}
\end{center}
\end{figure}

Most of the information about GRBs' host galaxies, however, comes from galaxies at $z < 1$. At larger distances, it becomes difficult to observe them even with today's largest telescopes. In fact, the majority of GRBs' host galaxies would probably remain unnoticed, but for the powerful GRBs that illuminate them. For a brief time, GRBs are background sources, and all the objects in the line of sight imprint their shadow signature in the form of absorption lines in GRBs' spectra. The spectroscopy of GRBs (their afterglows and host galaxies) is thus a very powerful diagnostic tool used not only for the determination of GRB distances, but also for the study of GRBs' environment and absorbers along the lines of sight:

\begin{itemize}
\item{{\bf Redshift distribution}:
thanks to their immense power, GRBs can be detected even at large redshifts. Fig. \ref{redshift} presents the GRBs' redshift distribution. Because the majority of GRB redshifts was determined using spectroscopic observations of optical afterglows and host galaxies, the sample is biased to optically bright GRBs. The current record holders are GRB~090423 with $z=8.26$ (spectroscopically determined) and GRB~090429B with even larger $z=9.4$ (determined photometrically, but not spectroscopically confirmed).

As GRBs can be detected to large distances, it would be very useful if their luminosities could be determined independently from their redshift distance. This would enable us to use them as standard candles to measure distances at even larger scales than with supernovae Type Ia, and to study the expansion of the universe. The best possibility seems to be the use of prompt gamma emission, which can be characterized by various parameters, such as duration, variability, lag, pulse rise/fall time, fluence, $E_{\rm iso}$ and $E_{\rm peak}$. Several correlations between these quantities have been found, but are still widely discussed.
}
\begin{figure}
\begin{center}
\includegraphics[angle=0, scale=1.0]{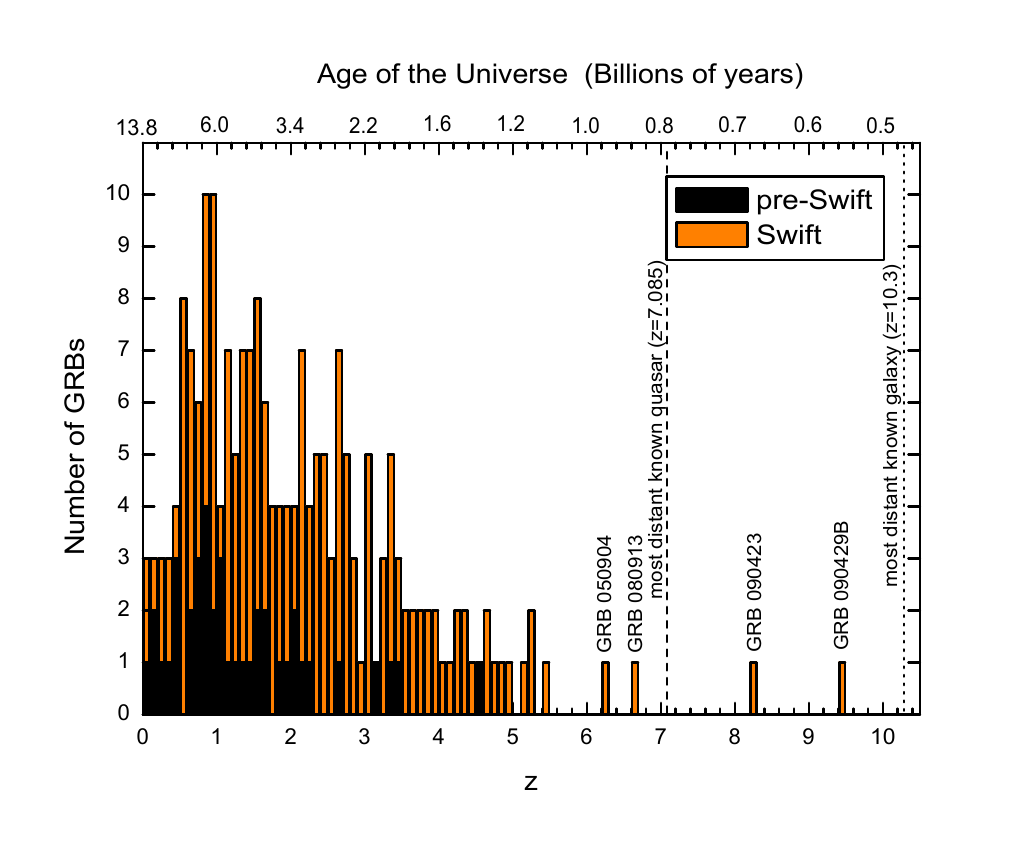}
\caption{GRBs redshift distribution before {\em Swift} and in {\em Swift} era. Vertical lines indicate redshifts of the most distant known quasar and galaxy (redshift of the latter has not been spectroscopically confirmed). Due to the higher sensitivity of {\em Swift} (in comparison to {\em BeppoSAX} and HETE-2), and follow-up campaigns, the redshift distributions of pre-{\em Swift} and {\em Swift} GRBs are different: average redshift of the {\em Swift} GRBs is larger:  $<z>\sim 2.1$ compared to  $<z>\sim 1.4$ before {\em Swift}.   
}
\label{redshift}
\end{center}
\end{figure}

\item{{\bf Metallicity of GRBs' host galaxies}:
the metallicity of GRBs' host galaxies is also a highly debated issue. Studies show that host galaxies of {\em long} GRBs have lower metallicity than field galaxies (i.e. galaxies that do not reside in galactic clusters) of the same mass. However, this does not necessarily mean that GRBs occur in special, low-metallicity galaxies, and that a direct link between low metallicity and production of GRBs exists. Indeed, there is a well-established link between the long GRBs and deaths of very massive stars, which produces a relation between long GRBs and star formation. It was shown that, in general, galaxies with a higher star formation rate have lower metallicities than more quiescent galaxies of the same mass, and a recent study \cite{mannucci} indicates this to hold also for the hosts of long GRBs. Recent studies \cite{mannucci, savaglio} are thus in agreement with the conclusion that GRBs' host galaxies are drawn from the normal population of star forming regions at $z<1$; nevertheless, it is wise to await more data and final answers on this topic. Since the localization of a host galaxy depends heavily on the detection of the optical afterglow, which can be extinguished by dust, it is possible that the current sample of host galaxies is biased against dust and high-metallicity galaxies. 
}
\item{{\bf Local environment}:  
by using spectroscopic observations of GRBs' afterglows, it is possible to study properties of the material surrounding GRBs. In high redshift GRBs, afterglow spectra show a wealth of absorption lines at the redshift of the GRB. The most pronounced is very strong, for most of the time damped, Lyman-$\alpha$ hydrogen absorption, which is produced by a large column density ($\log N ({\rm HI}) {\rm (cm^{-2})} > 20$) of neutral gas in the interstellar medium in the host galaxy. Bright quasars are also used to observe damped Lyman-$\alpha$ systems along the line of sight, but these observations are often biased towards regions of low density. Long GRBs, however, are associated with regions of star formation, and therefore are unique probes of the environment of high density gas and molecular clouds. 

Due to high column densities of neutral hydrogen, it is possible to determine chemical composition, detect low abundance elements, and also detect absorption lines from excited levels. The latter is particularly interesting, because GRBs' prompt and afterglow radiation is so intense that it can heat and photo-ionize material in its close vicinity, destroy molecules and vaporize dust grains. UV radiation can also change the population of different energy levels in atoms and molecules. Such UV pumping can be used to estimate the distance between the GRB and the absorber, since the closer the gas is to the GRB, the higher are the column densities of the excited levels. Results suggest that the power of a GRB affects a region of gas that is at least a few hundred light years in size \cite{petitjean}.
}
\begin{figure}
\begin{center}
\includegraphics[angle=0, scale=0.09]{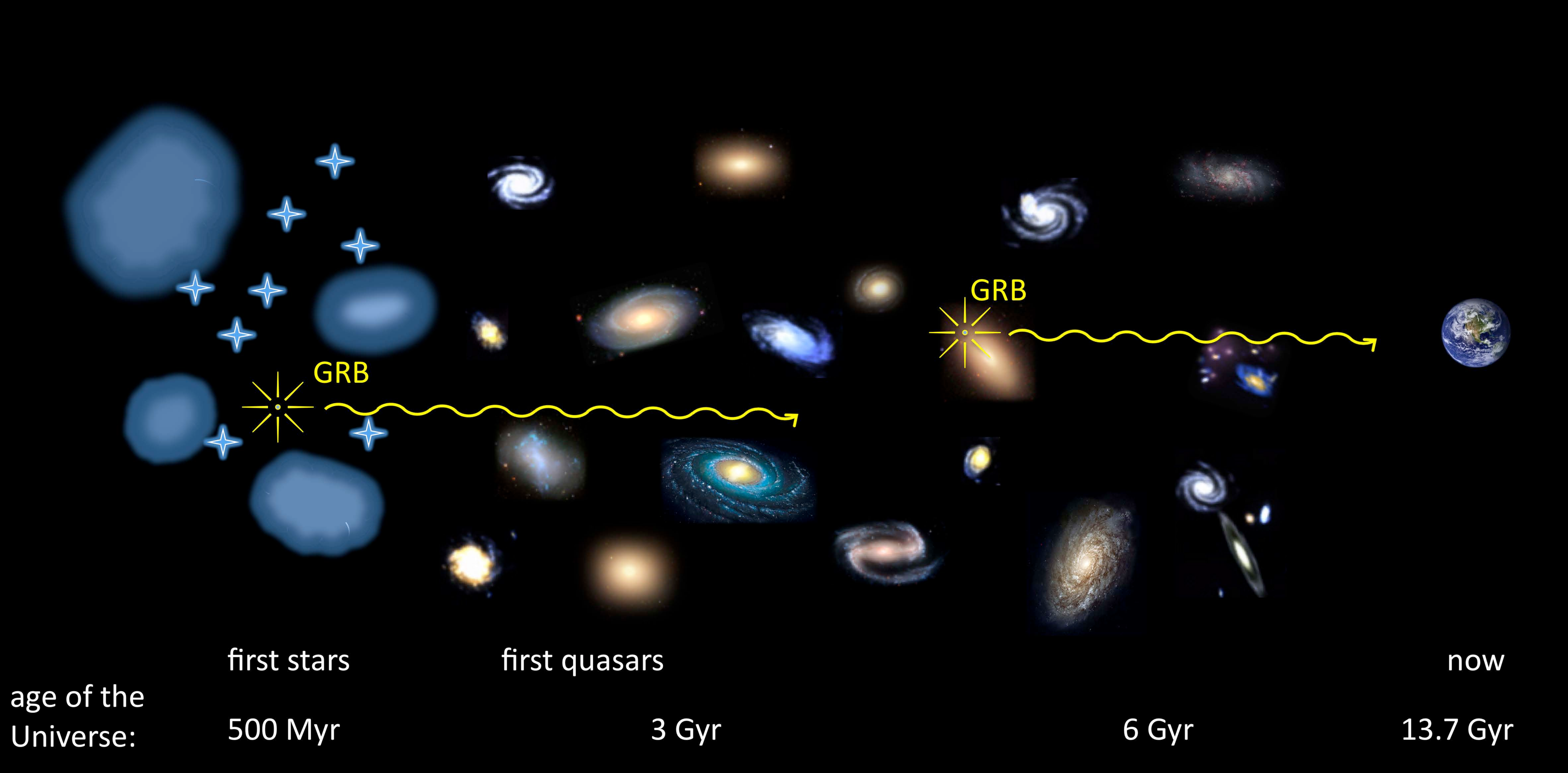}
\caption{Light from a GRB and its afterglows travels on its way to the Earth through circumburst medium, host galaxy medium and intervening absorbers. All of these may imprint their signature in the spectrum. From left to right: clouds of gas in the early universe collapsed to form the first (Pop III) massive stars, which probably produced the first GRBs. GRBs may have preceded the formation of the first galaxies and active galactic nuclei/quasars, which are powered by supermassive black holes and formed even later. Thus, GRBs may probe the properties and environment of the first stars and galaxies in the Universe, as well as properties of the intervening absorbers.}
\label{absorbers}
\end{center}
\end{figure}

 \item{{\bf Intervening absorbers}:
 light from GRBs can also be used for the study of intervening absorbers along GRBs' lines of sight. So far, observations show that the number of strong intervening Mg II absorbers\footnote[10]{As the name suggests, Mg II aborbers are gaseous structures that lie between the Earth and a distant object and give rise to Mg II absorption lines in the object's spectrum.} is a few times larger along GRBs' lines of sight than that measured for quasars' lines of sight over the same path length \cite{vergani}. This excess is not yet understood. 

}
\end{itemize}

\section{GRBs as cosmological probes}\label{cosmo-probes}
The extreme brightness of GRBs enables their detection out to distances of more than 13 billion light years. They are therefore indispensable as cosmological probes. They enable the study of end states of stellar evolution through the entire history of the universe: from relatively nearby bursts in the present universe to explosions of extremely massive young stars in the first galaxies in the early universe. Since GRBs are connected with star formation, they can be used to study cosmic star formation rate, which as far as we know today, could have been very different in the past. In addition, the light from GRBs illuminates distant galaxies and reveals properties of their local environment and of the material in the line of sight, thus making them unique cosmological tools for the study of the properties of high redshift galaxies and of the chemical enrichment of the universe. The discovery of GRBs at $z>8$ establishes that massive stars were being born, evolved and died as GRBs already less than $\sim 650$ million years after the Big Bang. GRBs are therefore of great importance for a better understanding of the early cosmic epochs during which the intergalactic medium was re-ionized by the radiation from the very first generation of massive stars and galaxies were in their first stages of formation.

\section{Conclusions}

GRBs are fascinating events, not only as the most powerful explosions known, but also as laboratories of stellar formation, end stages of stars, extreme relativistic physics, strong gravity regions, and objects probably endowed with strong magnetic fields. They can be observed throughout the Universe up to the first stars after the Big Bang. It is believed that their enormous electromagnetic output is only a fraction of the energy released, and that the majority of energy is carried away by neutrinos and gravitational waves. This makes GRBs ideal targets for multi-messenger astronomy projects. The voyage of discovering their secrets has been very exciting so far and it promises to continue bumpy and full of surprises in the future.

\section*{Acknowledgements}
AG acknowledges funding from the Slovenian Research Agency and from the Centre of Excellence for Space Sciences and Technologies SPACE-SI, an operation partly financed by the European Union, European Regional Development Fund and Republic of Slovenia, Ministry of Higher Education, Science and Technology.

\section*{Notes on contributor}
 
Andreja Gomboc obtained her PhD in Physics in 2001 at the Faculty of Mathematics and Physics, University of Ljubljana, Slovenia, working on simulations of tidal disruption of stars by supermassive black holes. As a post-doctoral Marie Curie Fellow at Astrophysics Research Institute, Liverpool John Moores University, UK (2002-2004), she started her research on GRBs, specifically on the rapid follow-up observations of GRBs optical afterglows with the robotic Liverpool Telescope at La Palma, Canary Islands. After her return to the University of Ljubljana, where she lectures in astronomy and theoretical astrophysics, she continues to be a member of the GRB group at Liverpool John Moores University, which received the Times Higher Award 2007 for research project of the year ({\em Measuring Gamma Ray Bursts}). She also collaborates with the GRB group at INAF Brera Astronomical Observatory, Italy and the Gaia Science Alerts Working Group, Cambridge, UK. In 2011, she became PI of the European Space Agency PECS project {\em Relativistic Global Navigation Satellite Systems}. Since the International Year of Astronomy 2009, she has been very active in public outreach, organizing astronomy exhibitions, public lectures and school competitions, and is the editor of the Slovenian version of the webpage Portal to the Universe (www.portavvesolje.si).

\medskip

\label{lastpage}

\end{document}